\documentclass[pdflatex,sn-mathphys-num,iicol]{sn-jnl}

\usepackage{graphicx}%
\usepackage{multirow}%
\usepackage{amsmath,amssymb,amsfonts}%
\usepackage{amsthm}%
\usepackage{mathrsfs}%
\usepackage[title]{appendix}%
\usepackage{xcolor}%
\usepackage{textcomp}%
\usepackage{manyfoot}%
\usepackage{booktabs}%
\usepackage{algorithm}%
\usepackage{algorithmicx}%
\usepackage{algpseudocode}%
\usepackage{listings}%

\theoremstyle{thmstyleone}%
\theoremstyle{thmstyletwo}%

\theoremstyle{thmstylethree}%

\begin{document}

\title{
Hadronic final-state effects on open-charm production at LHC energies}



\author[1]{\fnm{Taesoo} \sur{Song}}\email{T.Song@gsi.de}

\author*[1,2,3]{\fnm{Elena} \sur{Bratkovskaya}}\email{E.Bratkovskaya@gsi.de}

\author[4,5]{\fnm{Krzysztof} \sur{Redlich}}\email{Krzysztof.Redlich@uwr.edu.pl}


\affil*[1]{GSI Helmholtzzentrum f\"{u}r Schwerionenforschung GmbH, Planckstrasse 1, 64291 Darmstadt, Germany}

\affil[2]{Institute for Theoretical Physics, Johann Wolfgang Goethe Universit\"{a}t, Frankfurt am Main, Germany}
\affil[3]{Helmholtz Research Academy Hessen for FAIR (HFHF),GSI Helmholtz Center for Heavy Ion Research. Campus Frankfurt, 60438 Frankfurt, Germany}

\affil[4]{Institute of Theoretical Physics, University of Wroc\l aw, plac Maksa Borna 9, PL-50204 Wroc\l{}aw, Poland}
\affil[5]{Polish Academy of Sciences PAN, Podwale 75, PL-50449 Wroc\l{}aw, Poland}


\abstract{We investigate the influence of hadronic final-state interactions on open-charm observables in pp and Pb+Pb collisions at LHC energies within the microscopic Parton-Hadron-String Dynamics (PHSD) transport approach. Calculations with and without charm-hadron rescattering are compared with ALICE measurements and predictions of the Statistical Hadronization Model for charm. The relative integrated $D$-meson multiplicities are essentially insensitive to hadronic interactions. PHSD reproduces the weak system dependence of the nonstrange $D$-meson ratios and the increase of $D_s^+/D^0$ from pp to Pb+Pb collisions, consistent with the statistical-model trend from strangeness undersaturation towards equilibrium. In contrast, hadronic rescattering redistributes the transverse momentum spectra at low and intermediate momenta and enhances the elliptic flow. These results demonstrate the importance of accounting for the hadronic evolution when extracting QGP properties from differential open-charm observables.}




\maketitle

\section{Introduction}\label{sec1}


Charm quarks are considered to be sensitive probes of the properties of the strongly interacting matter produced in relativistic heavy-ion collisions. Due to their large masses, charm quarks are predominantly produced in the initial hard nucleon--nucleon collisions and subsequently interact with the partonic medium formed during the evolution of the collision. Their interactions with the quark--gluon plasma (QGP) modify their momentum distributions and generate collective flow, which are reflected in the transverse momentum spectra, nuclear modification factor $R_{AA}$, and elliptic flow $v_2$ of open-charm hadrons. However, after hadronization, charm hadrons may undergo further interactions with the surrounding hadronic medium. Thus, understanding the influence of hadronic final-state interactions is important for extracting reliable information on the properties of the QGP from experimental charm observables.

Recent measurements of $D^0$, $D^+$, $D^{*+}$, and $D_s^+$ meson production by the ALICE Collaboration in pp and Pb+Pb collisions at LHC energies provide detailed information on charm production and hadronization~\cite{ALICE:2019nxm,ALICE:2018lyv,ALICE:2021rxa,ALICE:2021kfc}. The Statistical Hadronization Model for charm (SHMc) assumes that charm quarks, produced in initial hard scatterings, hadronize statistically at the QCD phase boundary~\cite{Braun-Munzinger:2000csl,Braun-Munzinger:2000eyl,Andronic:2017pug,Andronic:2021erx,Andronic:2026cdu}. A recent analysis by Braun-Munzinger et al.~\cite{Braun-Munzinger:2024ybd} demonstrated that the relative yields of open-charm mesons exhibit remarkable regularities over a broad range of collision energies and system sizes. In particular, the $D^+/D^0$ and $D^{*+}/D^0$ ratios are found to be approximately independent of the collision system and charged-particle multiplicity, while the $D_s^+/D^0$ ratio reflects the influence of strangeness production.

A comparison of the statistical hadronization results with calculations based on a microscopic transport approach is particularly important to clarify the role of dynamical interactions and, especially, the influence of hadronic final-state rescattering on open-charm observables. Such a study requires a consistent microscopic description of the entire collision evolution, including both partonic and hadronic degrees of freedom and their interactions.

The Parton-Hadron-String Dynamics (PHSD) transport approach provides such an opportunity~\cite{Song:2015sfa,Song:2015ykw}. It describes the production and propagation of charm quarks in the QGP, their hadronization through coalescence and fragmentation, and the subsequent interactions of charm hadrons in the hadronic medium. Previous PHSD studies of Au+Au collisions at RHIC energies, $\sqrt{s_{NN}}=200$~GeV, demonstrated that hadronic rescattering affects the $D$-meson nuclear modification factor at low transverse momenta and enhances the elliptic flow $v_2$~\cite{Song:2015sfa}. The charm dynamics was further investigated within PHSD at LHC energies~\cite{Song:2015ykw}. Comparisons of the PHSD predictions with  ALICE measurements show that PHSD provides a good description of the $D$-meson nuclear modification factor and elliptic flow $v_2(p_T)$ as functions of transverse momentum~\cite{ALICE:2018lyv,ALICE:2017pbx}. These PHSD calculations include interactions during both the partonic and hadronic stages.

In the present study, we investigate the influence of hadronic final-state interactions on $D$-meson production in pp and Pb+Pb collisions at $\sqrt{s}=5.02, 7, 13$~TeV and $\sqrt{s_{NN}}=5.02$~TeV, respectively, within the PHSD approach. In particular, we explore the sensitivity of the relative $D$-meson multiplicities, transverse momentum spectra, and elliptic flow to hadronic rescattering by comparing calculations with and without final-state hadronic interactions. The results are confronted with ALICE measurements and predictions from the statistical hadronization model.

This paper is organized as follows. After the Introduction, we briefly review the description of charm production and interactions within PHSD in Section 2. In Section 3, we present the PHSD results for $D$-meson production in Pb+Pb and p+p collisions and compare them with SHM predictions and ALICE data. Finally, the conclusions are summarized in Sec.~\ref{summary}.

\section{Charm production and interactions in PHSD}\label{sec:model}

PHSD is a covariant microscopic transport approach that describes the evolution of relativistic heavy-ion collisions from the initial hard scatterings through the partonic phase and hadronization to the final hadronic interactions~\cite{Cassing:2009vt,Moreau:2019vhw}. 
The partonic medium is described by interacting off-shell quasiparticles within the Dynamical QuasiParticle Model \cite{Cassing:2007yg, Moreau:2019vhw, Soloveva:2019xph}, with properties constrained by lattice-QCD thermodynamics.

Charm production and propagation are implemented as described in Refs.~\cite{Song:2015sfa,Song:2015ykw}. 
Initial charm quarks are generated in hard nucleon--nucleon collisions using PYTHIA, with momentum distributions adjusted to perturbative-QCD calculations in Fixed-Order-Next-to-Leading-Logarithm (FONLL) \cite{Cacciari:2012ny}.
In the QGP, charm quarks undergo elastic scattering with light quarks, antiquarks, and gluons. During hadronization, they form charm hadrons through coalescence with nearby light quarks or fragmentation, after which the produced charm hadrons propagate and interact with the evolving hadronic medium.

The hadronic interactions of $D$ mesons with light mesons, including $\pi$, $K$, $\bar K$, and $\eta$, are described by unitarized effective amplitudes constrained by chiral and heavy-quark spin symmetries~\cite{Abreu:2011ic,Tolos:2013kva}. Scattering with baryons, such as nucleons and $\Delta$ resonances, is treated within a coupled-channel approach incorporating heavy-quark spin symmetry~\cite{Garcia-Recio:2008rjt,Romanets:2012hm,Tolos:2013kva}. These approaches account for dynamically generated resonances and provide energy-dependent cross sections for the microscopic transport calculation~\cite{Song:2015sfa}. In the present study, the influence of this stage is isolated by comparing PHSD calculations with and without charm-hadron rescattering, keeping the preceding partonic evolution and hadronization unchanged.

\section{Results}\label{sec2}

\begin{figure}[h]
\centerline{
\includegraphics[width=\columnwidth]{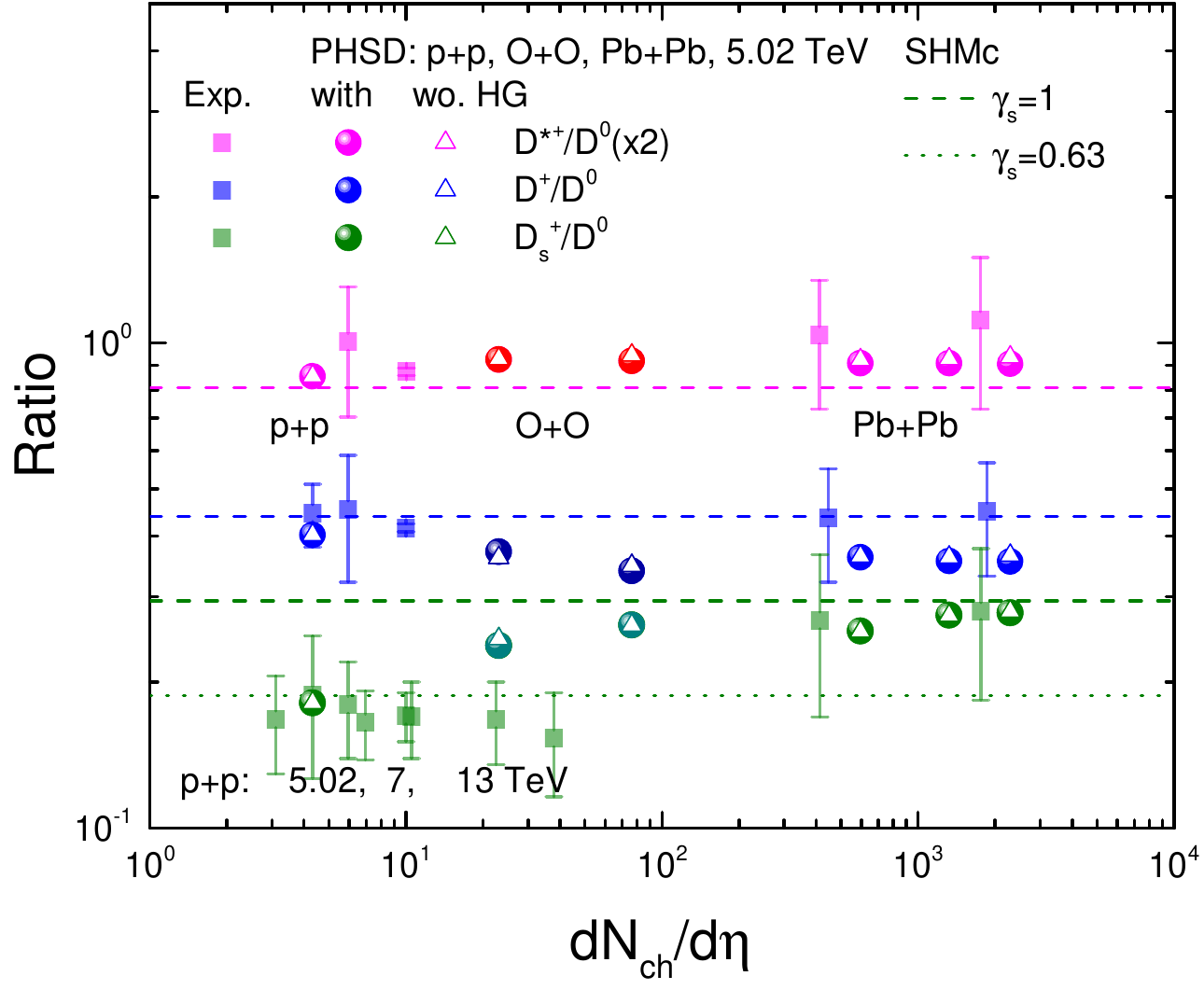}}
\caption{
Ratios of midrapidity $D$-meson multiplicities, $D^+/D^0$, $D^{*+}/D^0$, and $D_s^+/D^0$, as functions of the charged-particle pseudorapidity density $dN_{\rm ch}/d\eta$.
PHSD results are shown for pp collisions at $\sqrt{s}=5.02, 7$~ and 13 TeV and O+O and Pb+Pb collisions at $\sqrt{s_{NN}}=5.02$~TeV.
Solid circles denote calculations including hadronic rescattering, while open triangles correspond to calculations without hadronic rescattering.
Solid squares show the experimental data for pp collisions~\cite{ALICE:2019nxm,ALICE:2021mgk,ALICE:2017olh,ALICE:2023sgl,ALICE:2021npz} at $\sqrt{s}=5.02$, 7, and 13~TeV, together with $e^+e^-$ data at 91.25~GeV from LEP~\cite{Gladilin:2014tba}, and for Pb+Pb collisions~\cite{ALICE:2021rxa,ALICE:2021kfc}.
The dashed horizontal lines represent the SHMc predictions~\cite{Braun-Munzinger:2024ybd} at $T_c=156.5$~MeV with $\gamma_s=1$, while the dotted line corresponds to the $D_s^+/D^0$ ratio obtained with the strangeness undersaturation factor $\gamma_s=0.63$.
The $D^{*+}/D^0$ ratio is multiplied by two for visibility.
}
\label{RatioD}
\end{figure}

Figure~\ref{RatioD} shows the ratios of midrapidity $D$-meson multiplicities, $D^+/D^0$, $D^{*+}/D^0$, and $D_s^+/D^0$, as functions of the charged-particle pseudorapidity density. The PHSD results for pp, O+O, and Pb+Pb collisions are compared with ALICE measurements and SHMc predictions~\cite{Braun-Munzinger:2024ybd}. The nonstrange $D$-meson ratios exhibit only a weak dependence on the collision system and charged-particle multiplicity. PHSD reproduces the overall experimental trends, although its $D^+/D^0$ ratio is somewhat below the SHMc prediction, while $D^{*+}/D^0$ is slightly higher.

In contrast, the $D_s^+/D^0$ ratio increases significantly from pp to O+O collisions and approaches the values obtained in Pb+Pb already at moderate charged-particle multiplicities. This behavior can be explained by the increasing contribution of charm-quark coalescence relative to fragmentation with increasing system size. In PHSD, $D$ mesons are produced via two hadronization mechanisms: fragmentation of charm quarks and their coalescence with light or strange quarks from the partonic medium. In pp collisions, charm hadronization is dominated by fragmentation, while the coalescence contribution is very small. With increasing system size, the formation of a larger QGP volume enhances the probability of charm-quark coalescence. In particular, coalescence with strange quarks from the QGP favors $D_s$-meson production relative to nonstrange $D$ mesons, leading to an increase of the $D_s^+/D^0$ ratio. Thus, the O+O results reflect the growing importance of coalescence already in relatively small collision systems.

The PHSD result for pp collisions is close to the SHMc prediction with the strangeness undersaturation factor $\gamma_s=0.63$, whereas the O+O results lie between the SHMc predictions for $\gamma_s=0.63$ and $\gamma_s=1$ and approach the latter with increasing multiplicity. The Pb+Pb results are close to the equilibrium prediction with $\gamma_s=1$. Both PHSD and SHMc calculations are compatible with the available ALICE measurements within their uncertainties.

For all three ratios, the PHSD results with and without hadronic rescattering practically coincide. Thus, the relative $D$-meson abundances established during hadronization are essentially preserved throughout the subsequent hadronic evolution. The observed enhancement of $D_s^+/D^0$ with increasing system size is therefore predominantly driven by the increasing contribution of charm-quark coalescence in the partonic medium rather than by hadronic final-state interactions.

\begin{figure}[h]
\centerline{
\includegraphics[width=\columnwidth]{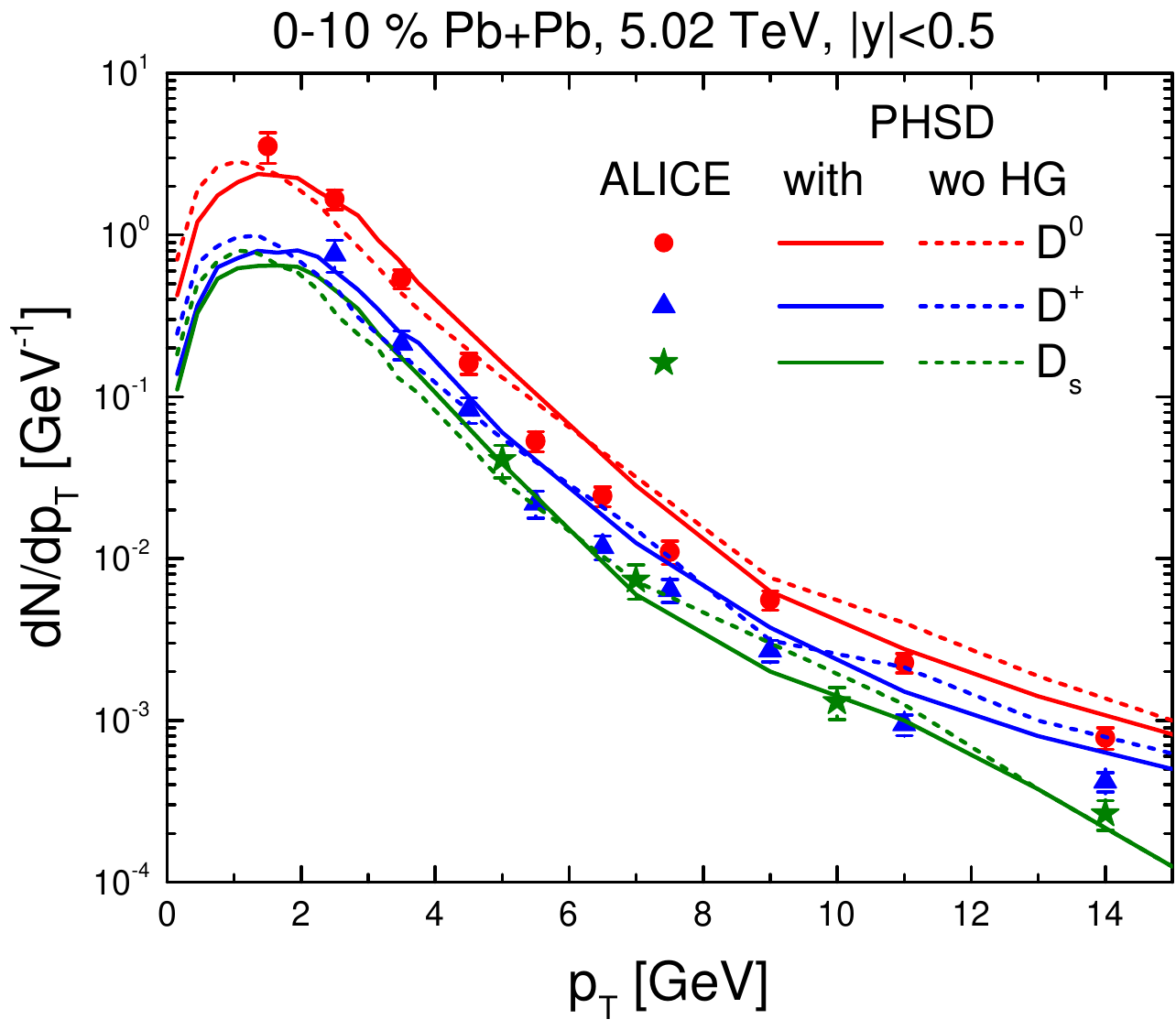}}
\caption{Transverse momentum distributions $dN/dp_T$ of prompt $D^0$ (red lines), $D^+$ (blue), $D_s$ (green) mesons in 0-10\% central Pb+Pb collisions at $\sqrt{s_{NN}}=5.02$~TeV at midrapidity $|y|<0.5$. 
Solid lines denote the PHSD calculations including hadronic interactions, while dashed lines correspond to calculations without hadronic rescattering.
Symbols show the ALICE data \cite{ALICE:2018lyv}.}
\label{ptD}
\end{figure}

The influence of hadronic interactions on momentum-dependent observables is illustrated in Fig.~\ref{ptD}, which displays the transverse momentum distributions of prompt $D^0$, $D^+$, and $D_s^+$ mesons in 0--10\% central Pb+Pb collisions at $\sqrt{s_{NN}}=5.02$~TeV.
The PHSD calculations provide an overall good description of the ALICE measurements~\cite{ALICE:2018lyv}. Hadronic interactions redistribute the yield in the low- and intermediate-$p_T$ region, lowering the low-$p_T$ peak and shifting strength towards somewhat larger transverse momenta. This reflects the additional momentum transfer to charm mesons through interactions with the expanding hadronic medium. At higher $p_T$, the calculations suggest a reduction of the yield, although the available statistics do not allow a firm conclusion in this region. Thus, hadronic rescattering modifies the shape of the spectra  while leaving the relative integrated $D$-meson multiplicities essentially unchanged.

\begin{figure}[h]
\centering
\includegraphics[width=1.1\columnwidth]{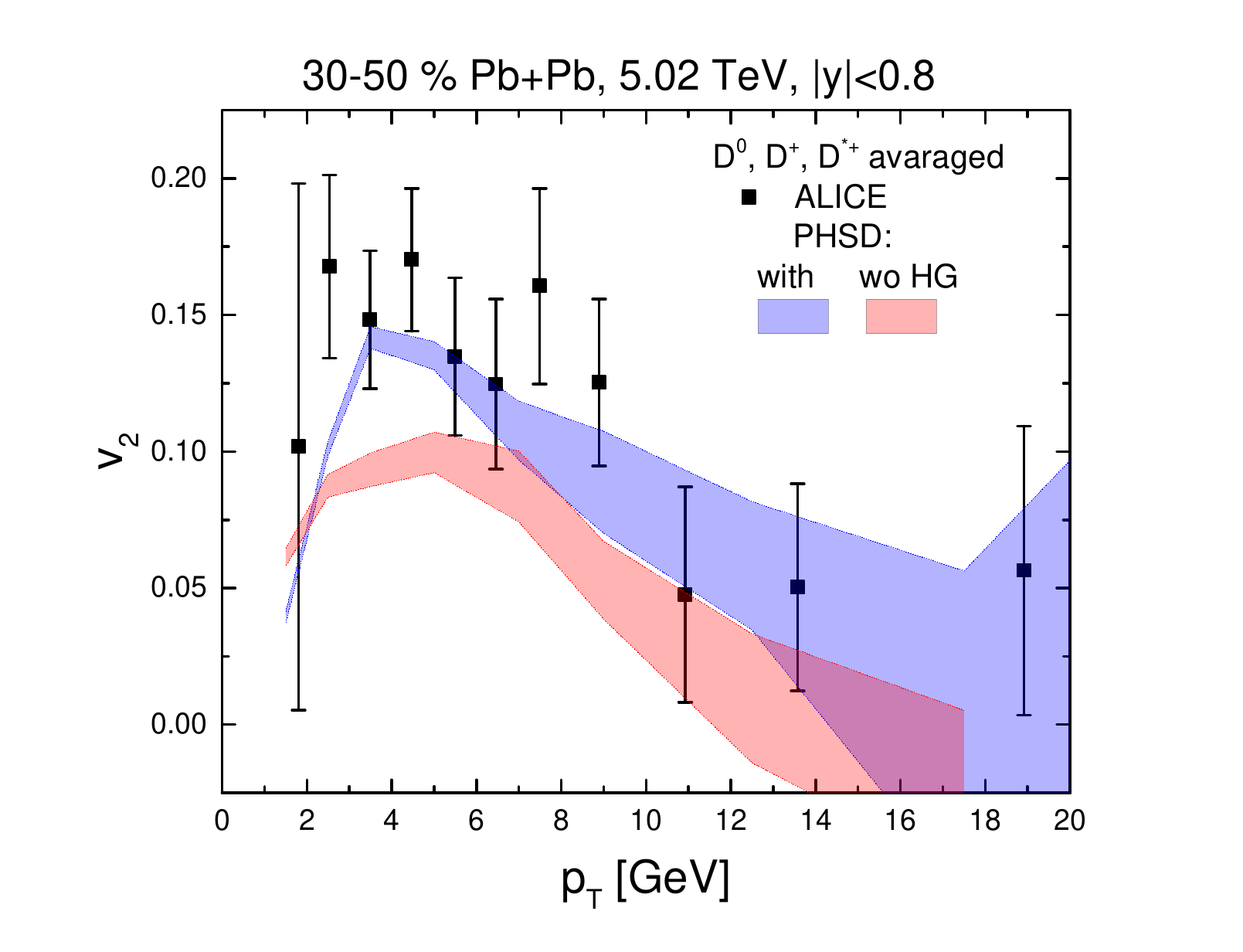}
\caption{Elliptic flow $v_2$ averaged over $D^0$, $D^+$, and $D^{*+}$ mesons in 30--50\% Pb+Pb collisions at $\sqrt{s_{NN}}=5.02$~TeV and midrapidity $|y|<0.8$. The blue and red bands show PHSD calculations with and without hadronic rescattering, respectively. The widths of the bands indicate the statistical uncertainties of the calculations. Solid squares show the ALICE measurements~\cite{ALICE:2017pbx}.}
\label{v2D}
\end{figure}

Figure~\ref{v2D} shows the elliptic flow averaged over $D^0$, $D^+$, and $D^{*+}$ mesons in 30--50\% Pb+Pb collisions. The blue and red bands represent the PHSD calculations with and without hadronic rescattering, respectively, including their statistical uncertainties. Hadronic rescattering enhances $v_2$ at low and intermediate transverse momenta, bringing the calculations closer to the ALICE measurements~\cite{ALICE:2017pbx}. This enhancement reflects the additional interaction of charm mesons with the collectively expanding hadronic medium. 

Taken together, these results show that hadronic final-state interactions have little influence on the relative integrated $D$-meson multiplicities, but modify the transverse momentum spectra and enhance the elliptic flow. Consequently, the hadronic contribution must be accounted for when using differential open-charm observables to constrain charm interactions in the QGP.

\section{Summary}\label{summary}

We have investigated the influence of hadronic final-state interactions on $D$-meson production in pp, O+O, and Pb+Pb collisions at LHC energies within the PHSD transport approach. The results have been compared with ALICE measurements and predictions of the Statistical Hadronization Model for charm.

The relative integrated $D$-meson multiplicities are essentially unaffected by hadronic rescattering. PHSD reproduces the weak system-size dependence of the nonstrange $D$-meson ratios and the increase of $D_s^+/D^0$ from pp to O+O and Pb+Pb collisions. This enhancement is driven by the increasing contribution of charm-quark coalescence relative to fragmentation with growing system size, which favors $D_s$-meson formation through recombination with strange quarks from the partonic medium. In particular, the O+O results already approach the Pb+Pb values at moderate charged-particle multiplicities. The PHSD results follow the trend of the statistical-model predictions from strangeness undersaturation in pp towards equilibrium in Pb+Pb collisions. Thus, the relative charm-meson abundances established during hadronization are largely preserved in the hadronic stage.

In contrast, hadronic interactions redistribute the transverse momentum spectra at low and intermediate $p_T$ and enhance the elliptic flow, improving the agreement with ALICE measurements. The comparison illustrates that similar integrated charm-meson ratios can emerge from statistical hadronization and microscopic transport, while differential observables retain sensitivity to the subsequent hadronic evolution. A consistent treatment of both partonic and hadronic interactions is therefore important for extracting QGP properties from open-charm measurements.

\bmhead{Acknowledgements}

We acknowledge support by the Deutsche Forschungsgemeinschaft (DFG, German Research Foundation) through the grant CRC-TR 211 'Strong-interaction matter under extreme conditions' - Project number 315477589 - TRR 211. 
The computational resources have been provided by the LOEWE-Center for Scientific Computing and the "Green Cube" at GSI, Darmstadt and by the Center for Scientific Computing (CSC) of the Goethe University. The publication is funded by the Open Access Publishing Fund of GSI Helmholtzzentrum fuer Schwerionenforschung. K.R. acknowledges the support from the National Science Centre (NCN), Poland, under OPUS Grant No. 2022/45/B/ST2/01527, and the support
of the Polish Ministry of Science and Higher Education.

\bibliography{EJPA_Charm_2026}


\begin{thebibliography}{27}
\ifx \bisbn   \undefined \def \bisbn  #1{ISBN #1}\fi
\ifx \binits  \undefined \def \binits#1{#1}\fi
\ifx \bauthor  \undefined \def \bauthor#1{#1}\fi
\ifx \batitle  \undefined \def \batitle#1{#1}\fi
\ifx \bjtitle  \undefined \def \bjtitle#1{#1}\fi
\ifx \bvolume  \undefined \def \bvolume#1{\textbf{#1}}\fi
\ifx \byear  \undefined \def \byear#1{#1}\fi
\ifx \bissue  \undefined \def \bissue#1{#1}\fi
\ifx \bfpage  \undefined \def \bfpage#1{#1}\fi
\ifx \blpage  \undefined \def \blpage #1{#1}\fi
\ifx \burl  \undefined \def \burl#1{\textsf{#1}}\fi
\ifx \doiurl  \undefined \def \doiurl#1{\url{https://doi.org/#1}}\fi
\ifx \betal  \undefined \def \betal{\textit{et al.}}\fi
\ifx \binstitute  \undefined \def \binstitute#1{#1}\fi
\ifx \binstitutionaled  \undefined \def \binstitutionaled#1{#1}\fi
\ifx \bctitle  \undefined \def \bctitle#1{#1}\fi
\ifx \beditor  \undefined \def \beditor#1{#1}\fi
\ifx \bpublisher  \undefined \def \bpublisher#1{#1}\fi
\ifx \bbtitle  \undefined \def \bbtitle#1{#1}\fi
\ifx \bedition  \undefined \def \bedition#1{#1}\fi
\ifx \bseriesno  \undefined \def \bseriesno#1{#1}\fi
\ifx \blocation  \undefined \def \blocation#1{#1}\fi
\ifx \bsertitle  \undefined \def \bsertitle#1{#1}\fi
\ifx \bsnm \undefined \def \bsnm#1{#1}\fi
\ifx \bsuffix \undefined \def \bsuffix#1{#1}\fi
\ifx \bparticle \undefined \def \bparticle#1{#1}\fi
\ifx \barticle \undefined \def \barticle#1{#1}\fi
\bibcommenthead
\ifx \bconfdate \undefined \def \bconfdate #1{#1}\fi
\ifx \botherref \undefined \def \botherref #1{#1}\fi
\ifx \url \undefined \def \url#1{\textsf{#1}}\fi
\ifx \bchapter \undefined \def \bchapter#1{#1}\fi
\ifx \bbook \undefined \def \bbook#1{#1}\fi
\ifx \bcomment \undefined \def \bcomment#1{#1}\fi
\ifx \oauthor \undefined \def \oauthor#1{#1}\fi
\ifx \citeauthoryear \undefined \def \citeauthoryear#1{#1}\fi
\ifx \endbibitem  \undefined \def \endbibitem {}\fi
\ifx \bconflocation  \undefined \def \bconflocation#1{#1}\fi
\ifx \arxivurl  \undefined \def \arxivurl#1{\textsf{#1}}\fi
\csname PreBibitemsHook\endcsname

\bibitem[\protect\citeauthoryear{Acharya et~al.}{2019}]{ALICE:2019nxm}
\begin{barticle}
\bauthor{\bsnm{Acharya}, \binits{S.}}, \betal:
\batitle{{Measurement of ${{\mathrm{D}}^0}$ , ${{\mathrm{D}}^+}$ ,
  ${{\mathrm{D}}^{*+}}$ and ${{\mathrm{D}}^+_{\mathrm{s}}}$ production in pp
  collisions at ${\sqrt{{\textit{s}}}~=~5.02~{\text {TeV}}}$ with ALICE}}.
\bjtitle{Eur. Phys. J. C}
\bvolume{79}(\bissue{5}),
\bfpage{388}
(\byear{2019})
\doiurl{10.1140/epjc/s10052-019-6873-6}
{\href{https://arxiv.org/abs/1901.07979}{{arXiv:1901.07979}}}
{[nucl-ex]}
\end{barticle}
\endbibitem

\bibitem[\protect\citeauthoryear{Acharya et~al.}{2018}]{ALICE:2018lyv}
\begin{barticle}
\bauthor{\bsnm{Acharya}, \binits{S.}}, \betal:
\batitle{{Measurement of D$^{0}$, D$^{+}$, D$^{*+}$ and D$_{s}^{+}$ production
  in Pb-Pb collisions at $ \sqrt{{\mathrm{s}}_{\mathrm{NN}}}=5.02 $ TeV}}.
\bjtitle{JHEP}
\bvolume{10},
\bfpage{174}
(\byear{2018})
\doiurl{10.1007/JHEP10(2018)174}
{\href{https://arxiv.org/abs/1804.09083}{{arXiv:1804.09083}}}
{[nucl-ex]}
\end{barticle}
\endbibitem

\bibitem[\protect\citeauthoryear{Acharya et~al.}{2022a}]{ALICE:2021rxa}
\begin{barticle}
\bauthor{\bsnm{Acharya}, \binits{S.}}, \betal:
\batitle{{Prompt D$^{0}$, D$^{+}$, and D$^{*+}$ production in Pb{\textendash}Pb
  collisions at $ \sqrt{s_{\mathrm{NN}}} $ = 5.02 TeV}}.
\bjtitle{JHEP}
\bvolume{01},
\bfpage{174}
(\byear{2022})
\doiurl{10.1007/JHEP01(2022)174}
{\href{https://arxiv.org/abs/2110.09420}{{arXiv:2110.09420}}}
{[nucl-ex]}
\end{barticle}
\endbibitem

\bibitem[\protect\citeauthoryear{Acharya et~al.}{2022b}]{ALICE:2021kfc}
\begin{barticle}
\bauthor{\bsnm{Acharya}, \binits{S.}}, \betal:
\batitle{{Measurement of prompt $D_s^+$-meson production and azimuthal
  anisotropy in Pb{\textendash}Pb collisions at $\sqrt {s_{NN}}$=5.02TeV}}.
\bjtitle{Phys. Lett. B}
\bvolume{827},
\bfpage{136986}
(\byear{2022})
\doiurl{10.1016/j.physletb.2022.136986}
{\href{https://arxiv.org/abs/2110.10006}{{arXiv:2110.10006}}}
{[nucl-ex]}
\end{barticle}
\endbibitem

\bibitem[\protect\citeauthoryear{Braun-Munzinger and
  Stachel}{2000}]{Braun-Munzinger:2000csl}
\begin{barticle}
\bauthor{\bsnm{Braun-Munzinger}, \binits{P.}},
\bauthor{\bsnm{Stachel}, \binits{J.}}:
\batitle{{(Non)thermal aspects of charmonium production and a new look at J /
  psi suppression}}.
\bjtitle{Phys. Lett. B}
\bvolume{490},
\bfpage{196}--\blpage{202}
(\byear{2000})
\doiurl{10.1016/S0370-2693(00)00991-6}
{\href{https://arxiv.org/abs/nucl-th/0007059}{{arXiv:nucl-th/0007059}}}
\end{barticle}
\endbibitem

\bibitem[\protect\citeauthoryear{Braun-Munzinger and
  Stachel}{2001}]{Braun-Munzinger:2000eyl}
\begin{barticle}
\bauthor{\bsnm{Braun-Munzinger}, \binits{P.}},
\bauthor{\bsnm{Stachel}, \binits{J.}}:
\batitle{{On charm production near the phase boundary}}.
\bjtitle{Nucl. Phys. A}
\bvolume{690},
\bfpage{119}--\blpage{126}
(\byear{2001})
\doiurl{10.1016/S0375-9474(01)00936-8}
{\href{https://arxiv.org/abs/nucl-th/0012064}{{arXiv:nucl-th/0012064}}}
\end{barticle}
\endbibitem

\bibitem[\protect\citeauthoryear{Andronic et~al.}{2018}]{Andronic:2017pug}
\begin{barticle}
\bauthor{\bsnm{Andronic}, \binits{A.}},
\bauthor{\bsnm{Braun-Munzinger}, \binits{P.}},
\bauthor{\bsnm{Redlich}, \binits{K.}},
\bauthor{\bsnm{Stachel}, \binits{J.}}:
\batitle{{Decoding the phase structure of QCD via particle production at high
  energy}}.
\bjtitle{Nature}
\bvolume{561}(\bissue{7723}),
\bfpage{321}--\blpage{330}
(\byear{2018})
\doiurl{10.1038/s41586-018-0491-6}
{\href{https://arxiv.org/abs/1710.09425}{{arXiv:1710.09425}}}
{[nucl-th]}
\end{barticle}
\endbibitem

\bibitem[\protect\citeauthoryear{Andronic et~al.}{2021}]{Andronic:2021erx}
\begin{barticle}
\bauthor{\bsnm{Andronic}, \binits{A.}},
\bauthor{\bsnm{Braun-Munzinger}, \binits{P.}},
\bauthor{\bsnm{K{\"o}hler}, \binits{M.K.}},
\bauthor{\bsnm{Mazeliauskas}, \binits{A.}},
\bauthor{\bsnm{Redlich}, \binits{K.}},
\bauthor{\bsnm{Stachel}, \binits{J.}},
\bauthor{\bsnm{Vislavicius}, \binits{V.}}:
\batitle{{The multiple-charm hierarchy in the statistical hadronization
  model}}.
\bjtitle{JHEP}
\bvolume{07},
\bfpage{035}
(\byear{2021})
\doiurl{10.1007/JHEP07(2021)035}
{\href{https://arxiv.org/abs/2104.12754}{{arXiv:2104.12754}}}
{[hep-ph]}
\end{barticle}
\endbibitem

\bibitem[\protect\citeauthoryear{Andronic et~al.}{2026}]{Andronic:2026cdu}
\begin{botherref}
\oauthor{\bsnm{Andronic}, \binits{A.}},
\oauthor{\bsnm{Braun-Munzinger}, \binits{P.}},
\oauthor{\bsnm{Redlich}, \binits{K.}},
\oauthor{\bsnm{Stachel}, \binits{J.}}:
{Experimental exploration of the QCD phase diagram}
(2026)
{\href{https://arxiv.org/abs/2606.14799}{{arXiv:2606.14799}}}
{[nucl-ex]}
\end{botherref}
\endbibitem

\bibitem[\protect\citeauthoryear{Braun-Munzinger
  et~al.}{2025}]{Braun-Munzinger:2024ybd}
\begin{barticle}
\bauthor{\bsnm{Braun-Munzinger}, \binits{P.}},
\bauthor{\bsnm{Redlich}, \binits{K.}},
\bauthor{\bsnm{Sharma}, \binits{N.}},
\bauthor{\bsnm{Stachel}, \binits{J.}}:
\batitle{{Emergence of new systematics for open charm production in high energy
  collisions}}.
\bjtitle{JHEP}
\bvolume{04},
\bfpage{058}
(\byear{2025})
\doiurl{10.1007/JHEP04(2025)058}
{\href{https://arxiv.org/abs/2408.07496}{{arXiv:2408.07496}}}
{[hep-ph]}
\end{barticle}
\endbibitem

\bibitem[\protect\citeauthoryear{Song et~al.}{2015}]{Song:2015sfa}
\begin{barticle}
\bauthor{\bsnm{Song}, \binits{T.}},
\bauthor{\bsnm{Berrehrah}, \binits{H.}},
\bauthor{\bsnm{Cabrera}, \binits{D.}},
\bauthor{\bsnm{Torres-Rincon}, \binits{J.M.}},
\bauthor{\bsnm{Tolos}, \binits{L.}},
\bauthor{\bsnm{Cassing}, \binits{W.}},
\bauthor{\bsnm{Bratkovskaya}, \binits{E.}}:
\batitle{{Tomography of the Quark-Gluon-Plasma by Charm Quarks}}.
\bjtitle{Phys. Rev. C}
\bvolume{92}(\bissue{1}),
\bfpage{014910}
(\byear{2015})
\doiurl{10.1103/PhysRevC.92.014910}
{\href{https://arxiv.org/abs/1503.03039}{{arXiv:1503.03039}}}
{[nucl-th]}
\end{barticle}
\endbibitem

\bibitem[\protect\citeauthoryear{Song et~al.}{2016}]{Song:2015ykw}
\begin{barticle}
\bauthor{\bsnm{Song}, \binits{T.}},
\bauthor{\bsnm{Berrehrah}, \binits{H.}},
\bauthor{\bsnm{Cabrera}, \binits{D.}},
\bauthor{\bsnm{Cassing}, \binits{W.}},
\bauthor{\bsnm{Bratkovskaya}, \binits{E.}}:
\batitle{{Charm production in Pb + Pb collisions at energies available at the
  CERN Large Hadron Collider}}.
\bjtitle{Phys. Rev. C}
\bvolume{93}(\bissue{3}),
\bfpage{034906}
(\byear{2016})
\doiurl{10.1103/PhysRevC.93.034906}
{\href{https://arxiv.org/abs/1512.00891}{{arXiv:1512.00891}}}
{[nucl-th]}
\end{barticle}
\endbibitem

\bibitem[\protect\citeauthoryear{Acharya et~al.}{2018}]{ALICE:2017pbx}
\begin{barticle}
\bauthor{\bsnm{Acharya}, \binits{S.}}, \betal:
\batitle{{$D$-meson azimuthal anisotropy in midcentral Pb-Pb collisions at
  $\mathbf{\sqrt{s_{\rm NN}}=5.02}$ TeV}}.
\bjtitle{Phys. Rev. Lett.}
\bvolume{120}(\bissue{10}),
\bfpage{102301}
(\byear{2018})
\doiurl{10.1103/PhysRevLett.120.102301}
{\href{https://arxiv.org/abs/1707.01005}{{arXiv:1707.01005}}}
{[nucl-ex]}
\end{barticle}
\endbibitem

\bibitem[\protect\citeauthoryear{Cassing and
  Bratkovskaya}{2009}]{Cassing:2009vt}
\begin{barticle}
\bauthor{\bsnm{Cassing}, \binits{W.}},
\bauthor{\bsnm{Bratkovskaya}, \binits{E.L.}}:
\batitle{{Parton-Hadron-String Dynamics: an off-shell transport approach for
  relativistic energies}}.
\bjtitle{Nucl. Phys. A}
\bvolume{831},
\bfpage{215}--\blpage{242}
(\byear{2009})
\doiurl{10.1016/j.nuclphysa.2009.09.007}
{\href{https://arxiv.org/abs/0907.5331}{{arXiv:0907.5331}}}
{[nucl-th]}
\end{barticle}
\endbibitem

\bibitem[\protect\citeauthoryear{Moreau et~al.}{2019}]{Moreau:2019vhw}
\begin{barticle}
\bauthor{\bsnm{Moreau}, \binits{P.}},
\bauthor{\bsnm{Soloveva}, \binits{O.}},
\bauthor{\bsnm{Oliva}, \binits{L.}},
\bauthor{\bsnm{Song}, \binits{T.}},
\bauthor{\bsnm{Cassing}, \binits{W.}},
\bauthor{\bsnm{Bratkovskaya}, \binits{E.}}:
\batitle{{Exploring the partonic phase at finite chemical potential within an
  extended off-shell transport approach}}.
\bjtitle{Phys. Rev. C}
\bvolume{100}(\bissue{1}),
\bfpage{014911}
(\byear{2019})
\doiurl{10.1103/PhysRevC.100.014911}
{\href{https://arxiv.org/abs/1903.10257}{{arXiv:1903.10257}}}
{[nucl-th]}
\end{barticle}
\endbibitem

\bibitem[\protect\citeauthoryear{Cassing}{2007}]{Cassing:2007yg}
\begin{barticle}
\bauthor{\bsnm{Cassing}, \binits{W.}}:
\batitle{{QCD thermodynamics and confinement from a dynamical quasiparticle
  point of view}}.
\bjtitle{Nucl. Phys. A}
\bvolume{791},
\bfpage{365}--\blpage{381}
(\byear{2007})
\doiurl{10.1016/j.nuclphysa.2007.04.015}
{\href{https://arxiv.org/abs/0704.1410}{{arXiv:0704.1410}}}
{[nucl-th]}
\end{barticle}
\endbibitem

\bibitem[\protect\citeauthoryear{Soloveva et~al.}{2020}]{Soloveva:2019xph}
\begin{barticle}
\bauthor{\bsnm{Soloveva}, \binits{O.}},
\bauthor{\bsnm{Moreau}, \binits{P.}},
\bauthor{\bsnm{Bratkovskaya}, \binits{E.}}:
\batitle{{Transport coefficients for the hot quark-gluon plasma at finite
  chemical potential $\mu_B$}}.
\bjtitle{Phys. Rev. C}
\bvolume{101}(\bissue{4}),
\bfpage{045203}
(\byear{2020})
\doiurl{10.1103/PhysRevC.101.045203}
{\href{https://arxiv.org/abs/1911.08547}{{arXiv:1911.08547}}}
{[nucl-th]}
\end{barticle}
\endbibitem

\bibitem[\protect\citeauthoryear{Cacciari et~al.}{2012}]{Cacciari:2012ny}
\begin{barticle}
\bauthor{\bsnm{Cacciari}, \binits{M.}},
\bauthor{\bsnm{Frixione}, \binits{S.}},
\bauthor{\bsnm{Houdeau}, \binits{N.}},
\bauthor{\bsnm{Mangano}, \binits{M.L.}},
\bauthor{\bsnm{Nason}, \binits{P.}},
\bauthor{\bsnm{Ridolfi}, \binits{G.}}:
\batitle{{Theoretical predictions for charm and bottom production at the LHC}}.
\bjtitle{JHEP}
\bvolume{10},
\bfpage{137}
(\byear{2012})
\doiurl{10.1007/JHEP10(2012)137}
{\href{https://arxiv.org/abs/1205.6344}{{arXiv:1205.6344}}}
{[hep-ph]}
\end{barticle}
\endbibitem

\bibitem[\protect\citeauthoryear{Abreu et~al.}{2011}]{Abreu:2011ic}
\begin{barticle}
\bauthor{\bsnm{Abreu}, \binits{L.M.}},
\bauthor{\bsnm{Cabrera}, \binits{D.}},
\bauthor{\bsnm{Llanes-Estrada}, \binits{F.J.}},
\bauthor{\bsnm{Torres-Rincon}, \binits{J.M.}}:
\batitle{{Charm diffusion in a pion gas implementing unitarity, chiral and
  heavy quark symmetries}}.
\bjtitle{Annals Phys.}
\bvolume{326},
\bfpage{2737}--\blpage{2772}
(\byear{2011})
\doiurl{10.1016/j.aop.2011.06.006}
{\href{https://arxiv.org/abs/1104.3815}{{arXiv:1104.3815}}}
{[hep-ph]}
\end{barticle}
\endbibitem

\bibitem[\protect\citeauthoryear{Tolos and Torres-Rincon}{2013}]{Tolos:2013kva}
\begin{barticle}
\bauthor{\bsnm{Tolos}, \binits{L.}},
\bauthor{\bsnm{Torres-Rincon}, \binits{J.M.}}:
\batitle{{D-meson propagation in hot dense matter}}.
\bjtitle{Phys. Rev. D}
\bvolume{88},
\bfpage{074019}
(\byear{2013})
\doiurl{10.1103/PhysRevD.88.074019}
{\href{https://arxiv.org/abs/1306.5426}{{arXiv:1306.5426}}}
{[hep-ph]}
\end{barticle}
\endbibitem

\bibitem[\protect\citeauthoryear{Garcia-Recio
  et~al.}{2009}]{Garcia-Recio:2008rjt}
\begin{barticle}
\bauthor{\bsnm{Garcia-Recio}, \binits{C.}},
\bauthor{\bsnm{Magas}, \binits{V.K.}},
\bauthor{\bsnm{Mizutani}, \binits{T.}},
\bauthor{\bsnm{Nieves}, \binits{J.}},
\bauthor{\bsnm{Ramos}, \binits{A.}},
\bauthor{\bsnm{Salcedo}, \binits{L.L.}},
\bauthor{\bsnm{Tolos}, \binits{L.}}:
\batitle{{The s-wave charmed baryon resonances from a coupled-channel approach
  with heavy quark symmetry}}.
\bjtitle{Phys. Rev. D}
\bvolume{79},
\bfpage{054004}
(\byear{2009})
\doiurl{10.1103/PhysRevD.79.054004}
{\href{https://arxiv.org/abs/0807.2969}{{arXiv:0807.2969}}}
{[hep-ph]}
\end{barticle}
\endbibitem

\bibitem[\protect\citeauthoryear{Romanets et~al.}{2012}]{Romanets:2012hm}
\begin{barticle}
\bauthor{\bsnm{Romanets}, \binits{O.}},
\bauthor{\bsnm{Tolos}, \binits{L.}},
\bauthor{\bsnm{Garcia-Recio}, \binits{C.}},
\bauthor{\bsnm{Nieves}, \binits{J.}},
\bauthor{\bsnm{Salcedo}, \binits{L.L.}},
\bauthor{\bsnm{Timmermans}, \binits{R.G.E.}}:
\batitle{{Charmed and strange baryon resonances with heavy-quark spin
  symmetry}}.
\bjtitle{Phys. Rev. D}
\bvolume{85},
\bfpage{114032}
(\byear{2012})
\doiurl{10.1103/PhysRevD.85.114032}
{\href{https://arxiv.org/abs/1202.2239}{{arXiv:1202.2239}}}
{[hep-ph]}
\end{barticle}
\endbibitem

\bibitem[\protect\citeauthoryear{Acharya et~al.}{2021}]{ALICE:2021mgk}
\begin{barticle}
\bauthor{\bsnm{Acharya}, \binits{S.}}, \betal:
\batitle{{Measurement of beauty and charm production in pp collisions at $
  \sqrt{s} $ = 5.02 TeV via non-prompt and prompt D mesons}}.
\bjtitle{JHEP}
\bvolume{05},
\bfpage{220}
(\byear{2021})
\doiurl{10.1007/JHEP05(2021)220}
{\href{https://arxiv.org/abs/2102.13601}{{arXiv:2102.13601}}}
{[nucl-ex]}
\end{barticle}
\endbibitem

\bibitem[\protect\citeauthoryear{Acharya et~al.}{2017}]{ALICE:2017olh}
\begin{barticle}
\bauthor{\bsnm{Acharya}, \binits{S.}}, \betal:
\batitle{{Measurement of D-meson production at mid-rapidity in pp collisions at
  ${\sqrt{s}=7}$ TeV}}.
\bjtitle{Eur. Phys. J. C}
\bvolume{77}(\bissue{8}),
\bfpage{550}
(\byear{2017})
\doiurl{10.1140/epjc/s10052-017-5090-4}
{\href{https://arxiv.org/abs/1702.00766}{{arXiv:1702.00766}}}
{[hep-ex]}
\end{barticle}
\endbibitem

\bibitem[\protect\citeauthoryear{Acharya et~al.}{2023}]{ALICE:2023sgl}
\begin{barticle}
\bauthor{\bsnm{Acharya}, \binits{S.}}, \betal:
\batitle{{Charm production and fragmentation fractions at midrapidity in pp
  collisions at $ \sqrt{\textrm{s}} $ = 13 TeV}}.
\bjtitle{JHEP}
\bvolume{12},
\bfpage{086}
(\byear{2023})
\doiurl{10.1007/JHEP12(2023)086}
{\href{https://arxiv.org/abs/2308.04877}{{arXiv:2308.04877}}}
{[hep-ex]}
\end{barticle}
\endbibitem

\bibitem[\protect\citeauthoryear{Acharya et~al.}{2022}]{ALICE:2021npz}
\begin{barticle}
\bauthor{\bsnm{Acharya}, \binits{S.}}, \betal:
\batitle{{Observation of a multiplicity dependence in the pT-differential charm
  baryon-to-meson ratios in proton{\textendash}proton collisions at s=13 TeV}}.
\bjtitle{Phys. Lett. B}
\bvolume{829},
\bfpage{137065}
(\byear{2022})
\doiurl{10.1016/j.physletb.2022.137065}
{\href{https://arxiv.org/abs/2111.11948}{{arXiv:2111.11948}}}
{[nucl-ex]}
\end{barticle}
\endbibitem

\bibitem[\protect\citeauthoryear{Gladilin}{2015}]{Gladilin:2014tba}
\begin{barticle}
\bauthor{\bsnm{Gladilin}, \binits{L.}}:
\batitle{{Fragmentation fractions of $c$ and $b$ quarks into charmed hadrons at
  LEP}}.
\bjtitle{Eur. Phys. J. C}
\bvolume{75}(\bissue{1}),
\bfpage{19}
(\byear{2015})
\doiurl{10.1140/epjc/s10052-014-3250-3}
{\href{https://arxiv.org/abs/1404.3888}{{arXiv:1404.3888}}}
{[hep-ex]}
\end{barticle}
\endbibitem

\end{thebibliography}
\end{document}